\documentclass[aps,prl,twocolumn,showpacs]{revtex4}
\usepackage{graphicx}

\begin{document}

\title{Multi-Band Exotic Superconductivity in the New Superconductor $Bi_4O_4S_3$}

\author{Sheng Li, Huan Yang, Jian Tao, Xiaxin Ding, and Hai-Hu Wen$^*$}

\affiliation{Center for Superconducting Physics and Materials,
National Laboratory of Solid State Microstructures and Department
of Physics, Nanjing University, Nanjing 210093, China}

\begin{abstract}
Resistivity, Hall effect and magnetization have been investigated
on the new superconductor $Bi_4O_4S_3$. A weak insulating behavior
has been induced in the normal state when the superconductivity is
suppressed. Hall effect measurements illustrate clearly a
multiband feature dominated by electron charge carriers, which is
further supported by the magnetoresistance data. Interestingly, a
kink appears on the temperature dependence of resistivity at about
4 K at all high magnetic fields when the bulk superconductivity is
completely suppressed. This kink can be well traced back to the
upper critical field $H_{c2}(T)$ in the low field region, and is
explained as the possible evidence of residual Cooper pairs on the
one dimensional chains.
\end{abstract}

\pacs{74.70.Dd, 74.20.Mn, 74.25.Dw, 74.25.fc}

\maketitle

Superconductivity is induced by the quantum condensation of large
number of paired electrons. The original proposal about the
pairing by the electron-phonon coupling is gradually replaced by
the more exotic pairing mechanism, such as through exchanging the
magnetic spin fluctuations, and the superconducting transition
temperature can be increased to a higher level. The
superconducting (SC) pairing mechanism of the
cuprates\cite{HTSMuller} and the iron pnictides\cite{Hosono},
although not yet settled down completely, should have a close
relationship with the electron correlations of the electrons from
the 3d orbits.\cite{PWAnderson,Scalapino,Canfield1} In the heavy
Fermion systems\cite{SiQM1} and organic materials\cite{Dressel}, a
similar conclusion may be achieved. In this regard, it is natural
to explore new superconductors with possible higher transition
temperatures in the compounds with transition metal elements. The
electrons in the p-orbital, especially those from the 5p or 6p
orbits, are normally assumed to have weak repulsive potential $U$
and quite wide band width $w$, and hence have very weak
correlation effect. It would be surprising to find out
superconductivity with exotic feature in the p-orbital based
compounds. Very recently, Mizuguchi et al.\cite{Mizuguchi1}
discovered superconductivity with $T_c^{onset}$ = 8.6 K and
$T_c^{zero}$ = 4.5 K in the so-called $BiS_2$ based compound
$Bi_4O_4S_3$. This compound has a layered structure with the space
group of $I4/mmm$ or $I-42m$. After just three days, the same
group reported superconductivity at about 10.6 K in another system
$LaO_{1-x}F_xBiS_2$ by doping electrons into the material through
substituting oxygen with fluorine.\cite{Mizuguchi2} Recalling that the same doping
technique was used by Kamihara et al.\cite{Hosono} in the
discovery of superconductivity in the FeAs-based superconductor
$LaO_{1-x}F_xFeAs$. Quickly followed is the theoretical work based
on the first principles band structure calculations,\cite{Kuroki}
which predicts that the dominating bands for the electron
conduction as well as for the superconductivity are derived from
the Bi 6$p_x$ and 6$p_y$ orbits. In this Letter, we present the
first set of data for an intensive study on the transport and
magnetic properties of the new superconductor $Bi_4O_4S_3$. Our
results clearly illustrate the multi-band, strong fluctuation and
some unexpected anomalous properties of superconductivity, putting
this interesting superconductor as an exotic one.

\begin{figure}
\includegraphics[width=9cm]{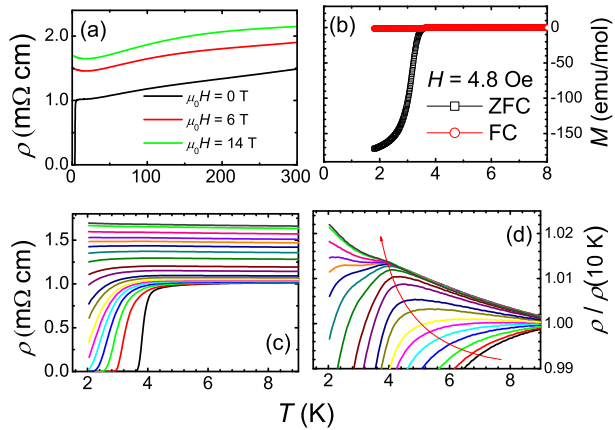}
\caption {(color online) (a) Temperature dependence of resistivity
at three magnetic fields: $\mu_0$H = 0, 6 and 14 T. It is clear
that, beside a moderate magnetoresistance effect, a weak
insulating behavior is induced by the magnetic field. (b) The
temperature dependence of magnetization near the superconducting
transition measured with H = 4.8 Oe in the field-cooled (FC) and
zero-field-cooled processes (ZFC). (c) An enlarged view for the
temperature dependence of resistivity at magnetic fields of (from
bottom to top) 0, 0.1 to 0.8 T with increments of 0.1 T; 1, 1.5, 2
to 7 with increments of 1 T; and 9, 12, 14 T. It is found that the
bulk superconductivity can be quickly suppressed by the magnetic
field, while the onset transition temperature changes slightly,
indicating a strong fluctuation effect. (d) Temperature dependence
of the resistivity (as shown in c) normalized at 10 K. A kink can
be clearly seen at about 4 K when the magnetic field is high and
the bulk superconductivity is suppressed completely. The red
arrowed line traces out the evolution from the superconducting
onset transition in the low field region to a kink at high
magnetic fields.} \label{fig1}
\end{figure}

\begin{figure}
\includegraphics[width=8cm]{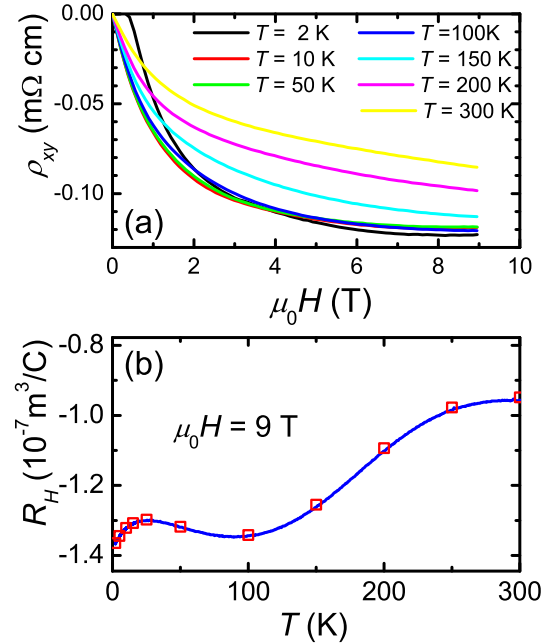}
\caption {(color online)(a) The transverse resistivity $\rho_{xy}$
versus the magnetic field $\mu_0H$ at different temperatures. The
curves are extremely curved, indicating a very strong multi-band
effect and/or the shallow band edge effect. (a) The Hall
coefficient $R_H$ calculated at 9 T, a clear temperature
dependence can be seen here, again supporting the multi-band
effect. The symbols represent the data measured by sweeping the
magnetic field at fixed temperatures. The solid line shows the
data measured at -9T and 9T by sweeping temperature. Both set of
data coincide very well.} \label{fig2}
\end{figure}

\begin{figure}
\includegraphics[width=8cm]{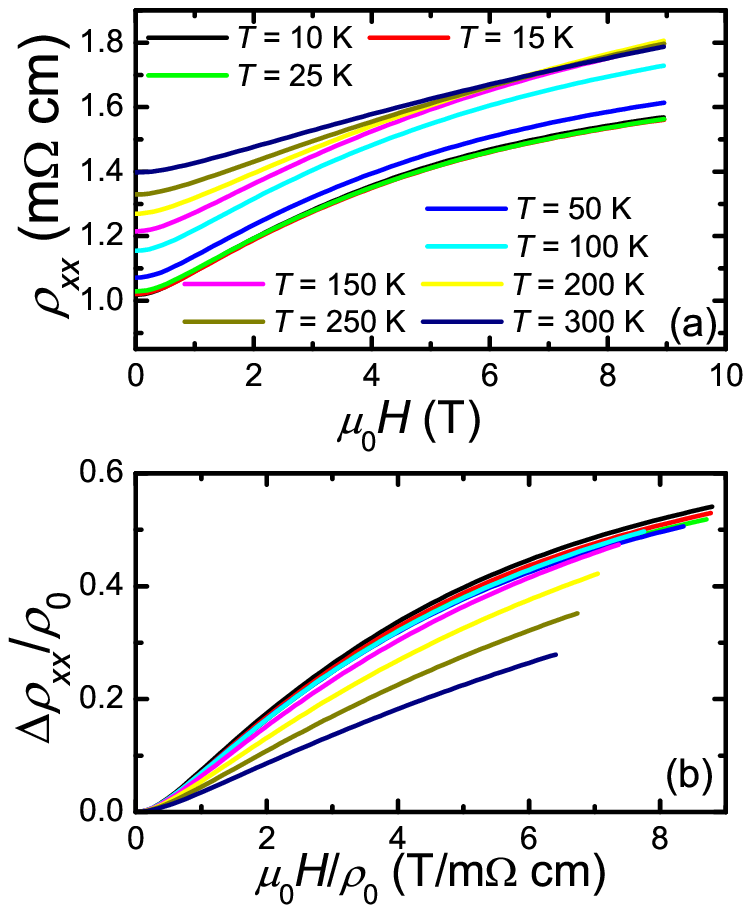}
\caption {(color online)(a) Field dependence of magnetoresistance
$\Delta\rho/\rho(0)$ at different temperatures. One can see that
the MR does decrease when the temperature is increased as expected by an electron-scattering. (b) Kohler plot for the
sample at different temperatures, and the Kohler's rule is not
obeyed. } \label{fig3}
\end{figure}

The polycrystalline samples were synthesized by using a two-step
solid state reaction method. First, the starting materials Bi
powder (purity 99.5\%, Alfa Aesar) and sulfa powders (purity
99.99\%, Alfa Aesar) were mixed in 2:3 ratio, ground and pressed
into a pellet shape. Then it was sealed in an evacuated quartz
tube and followed by annealing at 500$^\circ$C for 10 hours. The
resultant pellet was smashed and ground together with the
$Bi_2O_3$ powder (purity 99.5\%, Alfa Aesar) and sulfa powder, in
stoichiometry as the formula $Bi_4O_4S_3$. Again it was pressed
into a pellet and sealed in an evacuated quartz tube and burned at
510$^\circ$C for 10 hours. Then it was cooled down slowly to room
temperature. The second step was repeated in achieving good
homogeneity. The resultant sample looks black and very hard. We
cut the sample and obtained a specimen with a rectangular shape
(length 4.2 mm, width 2.0 mm and thickness of 0.125 mm).  Since it
is very hard, we can polish the sample with very fine sand paper
and obtained shiny and mirror like surface. The crystallinity of
the sample was checked by using the x-ray diffraction (XRD) with
the Brook Advanced D8 diffractometer with Cu K$_\alpha$ radiation.
The analysis of XRD data was done with the softwares Powder-X,
Fullprof and Topas. The XRD pattern looks very similar to that
reported by Mizuguchi et al.\cite{Mizuguchi1}. The Rietveld
fitting was done with the GSAS program, yielding a 90\% volume of
$Bi_4O_4S_3$ with 10\% of impurities which are mainly $Bi_2S_3$.
The resistivity, Hall effect were measured with Quantum Design
instrument PPMS-16T and PPMS-9T, and the magnetization was
detected by the Quantum Design instrument SQUID-VSM with a
resolution of about 5 $\times$ 10$^{-8}$ emu. The six-lead method
was used in the transport measurement on the longitudinal and the
transverse resistivity at the same time. The resistance and Hall
effect was measured by either sweeping magnetic field at a fixed
temperature or sweeping temperature at a fixed field, both set of
data coincide very well. The temperature stabilization was better
than $0.1\%$ and the resolution of the voltmeter was better than
10$\;$nV.

In Fig.~\ref{fig1}(a) we present the temperature dependence of resistivity
measured at three magnetic fields: $\mu_0H$ = 0, 6 and 14 T. In
addition to the moderate magnetoresistance, one can see that a
weak insulating behavior is induced by the magnetic field. This is
of course anti-intuitive for a normal state with Fermi liquid
characteristic. A simple explanation would be that the insulating
feature is given by an adjacent competing order here, once the
superconductivity is suppressed, the latter is promoted. However,
we should mention that the insulating behavior starts actually at
25 K (at 6 and 14 T) which is far beyond the superconducting
transition temperature here. Another possibility is that the
conduction band has a very shallow band edge, as illustrated by
the band structure calculations.\cite{Kuroki} When a magnetic
field is applied, the density of states (DOS) of the spin-up and
spin-down electrons will become asymmetric given by the Zeeman
effect. Therefore we have some polarized electrons which induce
the weak insulating behavior. In Fig.~\ref{fig1}(b) we present the
magnetization data measured in the zero-field-cooled (ZFC) and the
field-cooled mode (FC). The superconducting magnetic transition
starts at about 3.6 K, far below the onset transition temperature
which is about 10 K (determined by the deviating from the normal
state background of resistivity) and 4.6 K (determined in using
the crossing point of the normal state line and the extrapolation
of the steep transition line). Therefore strong superconducting
fluctuation is clearly present in this sample. This is actually
expected by the band structure calculations which predict a low
dimensionality of the electronic structure, and can be
corroborated by the quickly broadened transition under magnetic
fields, as shown in Fig.~\ref{fig1}(c). One can see that the bulk
superconductivity can be suppressed above 2 K by a magnetic field
as low as 0.9 T. However, even the bulk superconductivity can be
easily suppressed, a kink appears on the $\rho$ vs $T$ curve. If
following the onset transition of the resistivity, as shown in
Fig.~\ref{fig1}(d), we can clearly see that the kink can be traced back very
well with the upper critical field $H_{c2}(T)$ in the low field
region. Surprisingly, this kink stays at about 4 K even with a
magnetic field of 14 T. We interpret this kink as the residual
Cooper pairs existing in the system even the bulk
superconductivity is completely suppressed. Following the tendency
of this kink, a very high critical field can be expected in the
zero temperature limit, which certainly exceeds the Pauli limit
given by $\mu_0H_p = 1.84 T_c$ (T) (where $T_c$ is in unit of
Kelvin).\cite{PauliLimit}

In order to elucidate the normal state transport properties, we
did the Hall effect measurement. As shown in Fig.~\ref{fig2} (a),
the transverse resistivity remains negative at all temperatures,
indicating the electron like charge carriers as the dominating
one. It is remarkable that the field dependence of $\rho_{xy}$
exhibits an extremely curved feature at all temperatures,
prohibiting any single band description of Fermi liquid. Since the
plot $\rho_{xy}$ vs. $H$ is extremely curved, it is not possible
to use the slope of d$\rho_{xy}$/dT to determine the Hall
coefficient, we thus determine the Hall coefficient $R_H$ by using
the formula $R_\mathrm{H}=\rho_{xy}/H$ at a fixed magnetic field,
here 9T. The data is shown in Fig.~\ref{fig2} (b). A clear
temperature dependence of the Hall coefficient $R_H$ can be seen.
One can see that the temperature dependence of the Hall
coefficient $R_H$ is non-monotonic. An enhancement of $R_H$ is
observed below 25 K, which coincides very well with the starting
point of the insulating behavior as seen from the resistivity
data. Taking the value at 2 K, and using a single band assumption,
we get the charge carrier density of $4.5\times 10^{19}/cm^3$.
This indicates that the superconductor, as the cuprates and the
iron pnictides, has a very low superfluid density.

In the study of the Hall effect of clean $MgB_2$ superconductors,
our group has also observed a curved $\rho_{xy}$ vs. $H$ and a
strong temperature dependence of $R_H$, this was explained very
well by the multiband scattering effect.\cite{YangHall} As
explained in that paper, a natural consequence of the multiband
effect is that one should see a strong magnetoresistance effect.
It is well known that the magnetoresistance is a very powerful
tool to investigate the electronic scattering process and the
message of the Fermi surface. As mentioned above, in MgB$_2$, a
large magnetoresistance (MR) was found which is closely related to
the multiband property.\cite{LiQ,YangHall} For the present new
superconductor, we also measured and found a moderate MR, as shown
in Fig.~\ref{fig1}(a) and (c). In Fig.~\ref{fig3} (a) we present the field
dependence of the longitudinal resistivity. Clearly a 40\% to 50\%
increase of resistivity can be observed at a magnetic field of 9
T. In Fig.~\ref{fig3}(b) we present the MR ratio, i.e., $\Delta\rho/\rho_0$,
where $\rho_{xx}$ is the resistivity and $\rho_0$ is the
resistivity at zero field. One can see that the MR is about 52\%
at 10$\;$K and 9$\;$T. This MR ratio is significant and difficult
to be understood with a single band picture of the Fermi liquid.
Considering that the sample is a polycrystalline sample, the MR
effect may be weakened by mixing the transport components with the
magnetic field along different directions of the crystallographic
axes. For a single band metal with a symmetric Fermi surface, the
Kohler's law is normally obeyed. Kohler's law\cite{Kohler} shows
that the magnetoresistance $\Delta\rho_{xx}/\rho_0$ measured at
different temperatures should be scalable with the variable
H/$\rho_0$. For MgB$_2$, the Kohler's law is not obeyed because of
the multiband property.\cite{LiQ} We also do the scaling based on
the Kohler's law for this sample, the result is shown in
Fig.~\ref{fig3} (b). Clearly, the data measured at different
temperatures do not overlap and the Kohler's law is seriously
violated for this material. This discrepancy indicates that in
this new superconductor there is clear multiband effect, as
expected by the theoretical calculations.\cite{Kuroki} However,
the MR in this sample do show some specialty. For two-band or
multiband materials with weak MR, the MR ratios could be well
described by the expression $\Delta\rho/\rho_0\propto H^2$ in the
low field region. This is because the contribution of the higher
order even terms of $\mu_0H$ could be omitted at a low field (the
odd terms are absent here according to the Boltzmann equation for
electronic transport). But for the present material, the MR ratio
exhibits a roughly linear feature in the low field region,
deviating from the expectation. This effect is also found in
NbSe$_2$ which has a complex Fermi surface structure.\cite{NbSe2}
In this sense, the anomalies mentioned above in the present new
superconductor may also be induced by multiband effect together
with complex Fermi surfaces.

\begin{figure}
\includegraphics[width=9cm]{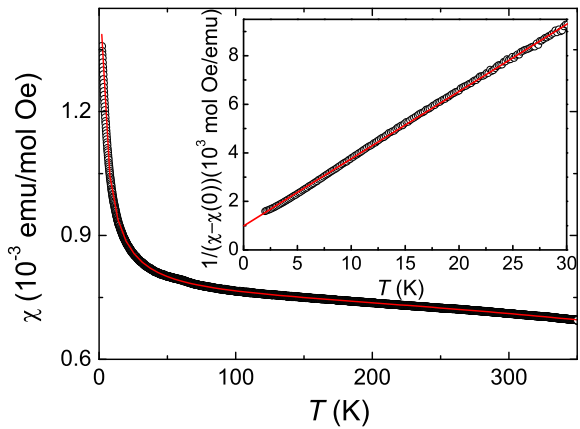}
\caption {(color online) Main panel: Temperature dependence of the
magnetic susceptibility measured at 3 T. The solid line is a fit
to the theoretical expression (see text). The inset gives a fit to
the data in low temperature region.} \label{fig4}
\end{figure}

Next we have a look at whether there is a sizable local magnetic
moment. For a system with the magnetic fluctuation given by the
local moments, the magnetization have two major origins: Pauli
susceptibility and the ionic (orbital and the nuclei)
contribution. Assuming the total magnetic susceptibility is given
by $\chi(T) = \chi(0)[1-({T}/{T_E})^2]+{C}/(T+T_0)$. The first
term comes from the Pauli susceptibility corrected with a
temperature dependence of the DOS at the Fermi energy. The $T_E$
is a parameter proportional to the Fermi energy. The second term
is related to the magnetism arising probably from the
contributions of the local moments. In Fig.~\ref{fig4} we show the
temperature dependence of the magnetic susceptibility measured at
a field of 3 Tesla. One can see a plateau of the magnetic
susceptibility above 100 K, which is mainly contributed by the
conduction electrons (the Pauli susceptibility). In the low
temperature region, one can clearly see a divergence of the
susceptibility. By fitting the data to above equation, we get
$\chi(0)$ = 0.000736 $emu/mol Oe$, $T_E$ = 1335 K, $C$ = 0.0034
$emu K/mol Oe$ and $T_0$ = 3.32 K. Once $C$ is determined, we can
get the magnetic moment given by each Bi atom. It turns out that
$\mu_{eff}$/Bi = 0.096$\mu_B$. This indicates that the local
moment of the Bi atom is very weak although with a finite value.

\begin{figure}
\includegraphics[width=9cm]{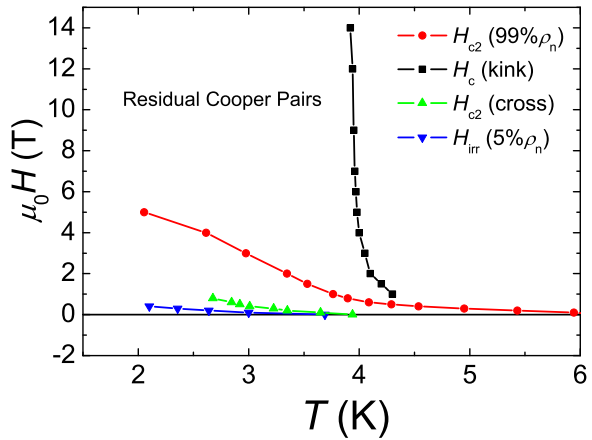}
\caption {(color online) Phase diagram derived from the resistive
transition curves. The onset transition point gives rise to the
upper critical field $H_{c2}$ shown by the red filled circles, the
bulk superconductivity is established in a very small area covered
by the curve name H$_{irr}$(T) (blue down triangles). The large
area between them indicates a strong superconducting phase
fluctuation. The curve marked with H$_{c2}$ (cross) marks the
upper critical field determined using the usual crossing point of
the normal state background and the extrapolated line of the steep
resistive transition part. The curve H$_c$ (kink) shows the point
determined from the kinky point of the resistive data shown in
Fig.~\ref{fig1}(d).} \label{fig5}
\end{figure}

Finally we present a phase diagram based on our transport and
magnetization measurement and give discussions on the possible
mechanism of superconductivity. As shown in Fig.~\ref{fig5}, the onset superconducting
transition point giving rise to the upper critical field $H_{c2}$
determined by $99\%\rho_n$ is shown by the red filled circles. The
bulk superconductivity is established in a very small area covered
by the irreversibility line H$_{irr}$(T) (blue down triangles).
The large area between them indicates a strong superconducting
phase fluctuation. This is actually consistent with the
theoretical expectation because the electronic system has an one
dimensional feature. The curve marked with H$_{2c}$ (cross) gives
the upper critical field determined using the usual crossing point
of the normal state background and the extrapolated line of the
steep resistive transition part. The most puzzling point is the
kink appearing in the $\rho$ vs. T data at a high magnetic field.
The curve marked with H$_c$(kink) shows the critical field
determined from the kinky point of the resistive data shown in
Fig.~\ref{fig1}(d), by following the trace of the arrowed red line there.
Since this line traces very well to the onset transition point
marked by H$_{c2}(99\%\rho_n)$ in the low field region, we
naturally attribute it to the existence of residual Cooper pairs.
If this kink can be interpreted as the onset for the pairing, that
would indicate a very strong pairing strength. In a simple BCS
argument, we have $H_{c2}=(\pi\Phi_0)/2\hbar v_F^2)\Delta_{sc}^2$,
where $\Phi_0$ is the flux quanta, $v_F$ is the Fermi velocity.
Such a strong pairing needs certainly a reasonable cause, which
exceeds the limit of the simple phonon mediated picture. Taking
account of the weak correlation effect in the Bi 6p electrons,
some other novel mechanism, such as the valence fluctuation of the
Bi$^{2+}$ and Bi$^{3+}$, may play an important role in this new
superconductor.

In summary, resistivity, Hall effect and magnetization have been
intensively investigated on the new superconductor $Bi_4O_4S_3$. A
weak insulating behavior is induced in the normal state when a
high magnetic field is applied. This can be induced either by an
adjacent competing order, or the very shallow $p_x$ and $p_y$ band
and small Fermi energy. Both the strong non-linear Hall effect and
the magnetoresistance indicate the multiband feature. A kink
appears on the temperature dependence of resistivity at all high
magnetic fields when the bulk superconductivity is completely
suppressed. This kink can be well traced back to the upper
critical field $H_{c2}(T)$ in the low field region. We argue that
the superconducting pairing, for some reason, could be very strong
and occur first in the one dimensional chains, then the bulk
superconductivity is established through the Josephson coupling
between them.

We appreciate the useful discussions with Qianghua Wang and
Jianxin Li. This work is supported by the NSF of China, the
Ministry of Science and Technology of China (973 projects:
2011CBA00102), and Chinese Academy of Sciences.

$^*$ hhwen@nju.edu.cn

\end{document}